\documentclass[a4paper,fleqn]{cas-sc}

\usepackage[numbers]{natbib}
\usepackage{mathrsfs}

\def\lmatrix{\left(\begin{array}}
\def\rmatrix{\end{array}\right)}
\def\bea{\begin{eqnarray}}
\def\eea{\end{eqnarray}}

\newcommand{\bZ}{\mathbb{Z}}

\begin{document}
\let\WriteBookmarks\relax
\def\floatpagepagefraction{1}
\def\textpagefraction{.001}

\shorttitle{QFT on flat, compact, orientable manifolds}

\shortauthors{Sebestyen Nagy, Daniel Nogradi}  

\title [mode = title]{QFT on flat, compact, orientable manifolds}  

\author{Sebestyen Nagy}

\ead{senbeastgyen@student.elte.hu}

\author{Daniel Nogradi}[orcid=0000-0002-3107-1958]

\ead{nogradi@bodri.elte.hu}

\affiliation{organization={Eotvos Lorand University, Department of Theoretical Physics},
            addressline={Pazmany Peter setany 1/a}, 
            city={Budapest},
            postcode={1117}, 
            country={Hungary}}

\begin{abstract}
    Motivated by the recent observation that rotating space-times at finite temperature can
    be described by particular flat, compact, orientable manifolds,
    which are different from $\mathbb T^4$, we catalogue all such 4-manifolds and their properties.
    There are in total 26 flat, compact, orientable 4-manifolds, 23 of
    them spin-manifolds, so fermions and a Dirac operator can also be defined on these. 
    Seven of them are appropriate for finite volume periodic boxes undergoing rotation at finite temperature.
\end{abstract}


\maketitle

\section{Introduction}
\label{introduction}

In \cite{Nagy:2025qvg} it was shown that a path integral description of a rotating spatial box at finite
temperature can be described by particular flat, compact, orientable 4-manifolds. The non-trivial topology
is a direct result of the rotation and comes about from a temporal boundary condition which is only
periodic up to a spatial rotation. These rotated boundary conditions mean that the fields are
well-defined on ${\mathbb T}^4 / {\mathbb Z}_k$, a smooth 4-manifold. There are 7 different topologies, 2
for $k=2,3,4$ and one for $k=6$. Rotating spatial volumes at finite temperature are under active 
study both on the lattice \cite{Yamamoto:2013zwa, Ambrus:2014uqa, Chernodub:2016kxh, Chernodub:2017ref, Braguta:2021jgn, Braguta:2023iyx, Braguta:2023yjn, Braguta:2023tqz}
and otherwise 
\cite{Vilenkin:1979ui,Erdmenger:2008rm,Banerjee:2008th,Son:2009tf,Kharzeev:2010gr,Landsteiner:2011cp,Kharzeev:2015znc,
Chen:2015hfc,Ebihara:2016fwa,Fukushima:2018grm,Fukushima:2020ncb,Chen:2022smf,Chernodub:2022veq,Chernodub:2022qlz,
Chen:2024tkr,Fukushima:2025hmh}, which was the main motivation for \cite{Nagy:2025qvg}.

The experimental motivation for all of these studies comes from scenarios where the rotation of QCD matter
cannot be ignored. For instance in fast rotating neutron stars 
\cite{Grenier:2015pya,Watts:2016uzu,Paschalidis:2016vmz} and in relativistic 
off-central heavy ion collisions \cite{Becattini:2007sr,Jiang:2016woz}. Of course, there are many orders
of magnitude differences in angular velocities between the two scenarios. In the latter case, which was
the primary focus of recent lattice studies, the experimental findings on $\Lambda$ hyperon spin polarizations indicate 
vorticity of the plasma may reach $\sim 10\, MeV$ \cite{STAR:2017ckg}.

The flat, compact, orientable 4-manifolds ${\mathbb T}^4 / {\mathbb Z}_k$ are actually only 7 out of 26
total cases (beyond the trivial $\mathbb T^4$). In this paper we catalogue the properties of all of them
and point out possible further physical applications for some of them.

\section{Bieberbach manifolds}
\label{bieberbachmanifolds}

Our motivation is to study Euclidean QFT on flat, compact, orientable 4-manifolds. These are locally
indistinguishable from ${\mathbb R}^4$ which is their main advantage over curved compact space-times 
such as $S^4$ or $S^1 \times S^3$. The prototypical example is ${\mathbb T}^4$, the 4-torus. It is a
well-known but non-trivial result that there are finitely many choices for the topology of
such spaces in any dimensions, in particular in 4-dimensions there are 26 choices beside the 4-torus
\cite{bieberbach1, bieberbach2}; see \cite{bbnwz, carat, book} for more details.
Collectively they are known as Bieberbach manifolds and can be characterized as ${\mathscr M} = {\mathbb T}^4 / G$
with a finite group $G$ acting on the 4-torus without fixed points and in such a way that 
${\mathscr M}$ is a smooth 4-manifold.

Let us start from ${\mathbb R}^4$ with its standard metric. Introduce a lattice $\Lambda$ defined by four
basis vectors $e^a$ which collectively can be assembled into a $4\times 4$ matrix $e$ with $e^a$ the
$a^{th}$ column. The lattice $\Lambda$ consists of all integer linear combinations of the basis vectors
$e^a$,
\bea
\label{lat}
\Lambda = \{ n_a e^a \; |\; n_a \in {\mathbb Z} \} = {\mathbb Z}^4 \;.
\eea
Then ${\mathbb R}^4 / \Lambda = {\mathbb T}^4$ and the standard flat metric on $\mathbb R^4$ descends to
a flat metric on $\mathbb T^4$, which further descends to a flat metric on $\mathscr M$.
It is possible to realize ${\mathscr M}$ also as the
factor ${\mathscr M} = {\mathbb R}^4 / \Gamma$ where $\Gamma$, a Bieberbach group, acts by isometries. 
It is a subgroup of the Euclidean group $E(4)$ generated by $\Lambda$ and $G$.
The equivalence of the two realizations, ${\mathbb T}^4 / G = {\mathbb R}^4 / \Gamma$ can easily be understood by
considering that $\Gamma / \Lambda = G$, i.e. $\Gamma$ consists of rotations and translations, up to
translations by the lattice $\Lambda$. Hence the action of $\Gamma$ on $\mathbb R^4$ descends to an action
of $G$ on $\mathbb T^4$. 

The construction is, of course, only well-defined if $G$ leaves the lattice $\Lambda$ invariant and
there are no fixed points on ${\mathbb T}^4$. The first condition forces $G$ to be a finite subgroup
and as such there are only finitely many choices up to isomorphism and the second condition forces
$G$ to have non-trivial translations. If elements $g \in G$ are labelled
as $g = (A,b)$ with $A \in SO(4)$ and $b$ a translation, then the $A$ matrices form a finite subgroup of
$SO(4)$. The group $G$ preserving the lattice means that in the basis $e^a$ the matrix elements of
$A$ need to be an integer and the components of $b$ need to be rational,
\bea
\label{nr}
A e^{c} = N_{ac} e^{a} \qquad \qquad b = r_a e^a\;,
\eea
where $N_{ac} \in \mathbb Z$ and $r_a = k_a / |G|$ with $k_a \in \mathbb Z$. Similarly, let us label
elements of $\Gamma$ as $(A,b,e^{(n)})$ where $(A,b) \in G$ and $e^{(n)} = n_a e^a \in \Lambda$
where $n_a \in \mathbb Z$.

These conditions allow the following finite subgroups of $SO(4)$
to appear: ${\mathbb Z}_{2,3,4,6}$, ${\mathbb Z}_2
\times {\mathbb Z}_2$ which are all abelian, and the non-abelian groups $S_3, A_4, D_4, D_6$, 
using standard notation for the symmetric, dihedral and alternating groups. The number of generators is
either one or two and the corresponding $(N,r)$ form, so the $(A,b)$ form in the basis $e^a$, 
is listed for each case in appendix \ref{generators}. Although these are only 9 different groups $G$
and several $\Gamma$ share the same $G$ factor, the full groups $\Gamma$ in the 26 cases are all
different thanks to the different ways $G$ acts on the translational lattice part $\Lambda$.

Starting from a 3-dimensional construction
following the same line of reasoning as above, but with ${\mathbb T}^3$ and $G_3$, we obtain 3-dimensional
flat, compact, orientable spaces ${\mathscr M}_3 = {\mathbb T}^3 / G_3$. Then we can always construct
${\mathscr M} = {\mathscr M}_3 \times S^1$ which will of course again be flat, compact, orientable and
4-dimensional. The result of the 3-dimensional classification, known from classic crystallography, is
that there are 5 non-trivial topologically different options, so 5 out of the 26 4-dimensional spaces
are of this factorized type. The allowed groups in 3-dimensions are ${\mathbb Z}_{2,3,4,6}$ and ${\mathbb
Z}_2 \times {\mathbb Z}_2$. The appearance of non-abelian groups is thus a new feature in 4 or more dimensions.

\begin{table}
\begin{center}
\renewcommand{\arraystretch}{1.2}
\begin{tabular}{cccccc}
\hline
CARAT & $G$ & 3D & spin & $k$ & $n$ \\
\hline
min.18.1.1.1 & $\bZ_2$ & yes & yes & 2  & 6 \\
min.18.1.2.1 & $\bZ_2$ & no & yes & 2 & 6 \\
min.35.1.1.1 & $\bZ_3$ & yes & yes & 3 & 4 \\
min.35.1.2.1 & $\bZ_3$ & no & yes & 3 & 4 \\
min.25.1.1.1 & $\bZ_4$ & yes & yes & 4 & 4 \\
min.25.1.2.2 & $\bZ_4$ & no & yes & 4 & 4 \\
group.70.1.1.1 & $\bZ_6$ & yes & yes & 6 & 4 \\
min.22.1.1.11 & $\bZ_2 \times \bZ_2$ & yes & yes & --- & 4 \\
min.22.1.1.12 & $\bZ_2 \times \bZ_2$ & no & yes & 2 & 4 \\
min.22.1.1.13 & $\bZ_2 \times \bZ_2$ & no & yes & 2 & 4 \\
min.22.1.1.15 & $\bZ_2 \times \bZ_2$ & no & yes & --- & 4 \\
min.22.1.2.7 & $\bZ_2 \times \bZ_2$ & no & yes & 2 & 4 \\
min.22.1.4.7 & $\bZ_2 \times \bZ_2$ & no & no & --- & 4 \\
min.22.1.5.3 & $\bZ_2 \times \bZ_2$ & no & yes & --- & 4 \\
min.22.1.8.7 & $\bZ_2 \times \bZ_2$ & no & no & --- &  4 \\
min.22.1.12.6 & $\bZ_2 \times \bZ_2$ & no & yes & --- & 4 \\
min.38.1.1.1 & $S_3$ & no & yes & 2 &  3 \\
min.38.2.1.1 & $S_3$ & no & yes & 2 &  3 \\
min.38.2.3.1 & $S_3$ & no & yes & --- & 3 \\
min.29.1.1.13 & $D_4$ & no & yes & 2 & 3 \\
min.29.1.1.14 & $D_4$ & no & yes & 2 & 3 \\
min.29.1.1.15 & $D_4$ & no & yes & 2 & 3 \\
min.29.1.2.7 & $D_4$ & no & no & --- & 3 \\
group.81.1.1.5 & $D_6$ & no & yes & 2 & 3 \\
min.44.1.2.1 & $A_4$ & no & yes & --- & 2 \\
min.44.2.1.1 & $A_4$ & no & yes & 3 &   2 \\
\hline
\end{tabular}
\end{center}
    \caption{Summary of the 26 non-trivial flat, compact, orientable
    4-manifolds. The column $k$ indicates if a rotating space-time interpretation is 
    possible with $\vartheta = 2\pi/k$ or a dash otherwise.
    The last column $n$ is the number of free parameters of
    the compatible flat metrics; see text for details.}
    \label{table}
\end{table}

The topology of ${\mathscr M}$ is completely fixed by $\Gamma$. In particular the fundamental group is
$\pi_1(\mathscr M) = \Gamma$ which generalizes the 4-torus case $\pi_1(\mathbb T^4) = \mathbb Z^4$
where $G$ is just the identity element. 

The flat metric on ${\mathscr M}$ is also completely fixed by the standard flat metric on $\mathbb R^4$
since all group actions are isometries. The volume of ${\mathbb T}^4$ and ${\mathscr M}$ is 
$|\det(e)|$ and $|\det(e)|\, /\, |G|$, respectively.

The 26 groups are listed in Table \ref{table}, labelled by their standard CARAT name from
crystallography \cite{carat}. Whether they are of the ${\mathscr M}_3 \times S^1$ form and whether they are spin
manifolds \cite{math1, math2, math3} are also listed. In \cite{Nagy:2025qvg} the notation for the seven
$G = {\mathbb Z}_k$ cases was simply, 
$\mathbb Z_2, \mathbb Z_2^\prime,\mathbb Z_3, \mathbb Z_3^\prime,\mathbb Z_4, \mathbb
Z_4^\prime,\mathbb Z_6$, which correspond to 
min.18.1.1.1, min.18.1.2.1, min.35.1.1.1, min.35.1.2.1, min.25.1.1.1, min.25.1.2.2, group.70.1.1.1,
respectively.

Once the lattice $\Lambda$ is given, the choice of basis vectors is of course not unique. Transformations
of the type,
\bea
\label{inv}
e \to R\, e\, K^{-1}\;,
\eea
where $R\in SO(4)$ and $K \in GL(4,{\mathbb Z})$ correspond to the same $\Lambda$ and hence the 
same ${\mathbb T}^4$. Here
the freedom with respect to $R$ is just an over-all orientation of all 4 basis vectors $e^a$, while $K$
corresponds to a relabeling of the integer vectors $n_a$.

\section{Compatible metrics}
\label{compatiblemetrics}

The previous section classified the 4-dimensional compact, flat, orientable manifolds topologically. 
For a QFT construction a metric is also needed, which descends from ${\mathbb R}^4$ to ${\mathbb
T}^4$ and then to ${\mathscr M} = {\mathbb T}^4 / G$. We start with the standard flat metric on ${\mathbb
R}^4$ and the metric on ${\mathbb T}^4$ is specified by the vierbein $e$. In coordinates $x = x_a e^a$
with $0 \leq x_a \leq 1$, the flat metric is determined by the scalar products of the basis vectors,
\bea
\eta_{ab} = e^a \cdot e^b \;.
\eea
For a given $G$, the choice of the vierbein and correspondingly the flat metric
$\eta_{ab}$ is not arbitrary, it should be compatible with the action of $G$, in particular the
rotational part of it. More precisely, expressing $(A,b) \in G$ in the basis of $e^a$, what we denoted by
$(N,r)$, we have,
\bea
N^T \eta N = \eta\;.
\eea
Hence the number of free parameters in the basis vectors, up to transformations (\ref{inv}), is given by
all the possible solutions of the above constraint for all $N$ integer rotational parts of $G$. This is
useful to know because it will determine what angles between the basis vectors are allowed for the
original ${\mathbb T}^4$ for a given $G$ such that ${\mathbb T}^4 / G$ is well-defined.
The number of free parameters, or in other words the dimension of the $G$-invariant metrics can
be calculated as
\bea
n = \frac{1}{|G|} \sum_{(A,b)\in G} \frac{\chi(A)^2 + \chi(A^2)}{2}\;,
\eea
where $\chi(A)$ is the character in the fundamental representation of $SO(4)$. Equivalently, if $\mathbb
R^4$ is decomposed into real irreducible representations of $G$ with multiplicities $n_i$, then,
\bea
n = \sum_i \frac{n_i (n_i + 1)}{2}\;.
\eea
In \cite{Nagy:2025qvg} the most general metrics and hence possible angles between basis vectors were
worked out explicitly for the abelian $G = \mathbb Z_k$ cases. The number of free parameters $z_i$ there, 
the overall scale $L$, the temporal
direction $L_0$ and the possible shift labelled by $y_3$ account for the total $n$ given
above. Hence the number of explicit continuous parameters in the metrics, the $z_i$ variables, is $n-3$.  
We have $n=6,4,4,4$ for $k=2,3,4,6$ meaning that the number of free $z_i$ parameters
in the metrics are $3,1,1,1$, respectively, precisely matching \cite{Nagy:2025qvg}.
The values $n$ for all cases are listed as the last column in Table \ref{table}.

\section{Boundary conditions}
\label{boundaryconditions}

Functions on the spaces ${\mathscr M} = {\mathbb T}^4 / G$ can be realized in practice as follows.
Clearly, $G$-invariant functions on ${\mathbb T}^4$ are well-defined on ${\mathscr M}$.
Another equivalent option is to consider an elementary cell in ${\mathbb T}^4$, whose volume is
$vol({\mathbb T}^4) / |G|$, such that by applying $G$ we get the full ${\mathbb T}^4$, and define the
functions on this elementary cell only, subject to boundary conditions dictated by $G$. 

If there is only one generator we have $G = {\mathbb Z}_k$. The generator $g=(A,b)$ is such that $g^k =
1$ (up to lattice translations) and the elementary cell can be chosen as follows. 
The 4-torus is split into $k$ pieces along the shift
direction $b$ and one may choose any one of the $k$ subvolumes. Then a function $\phi(x)$ defined on this
elementary cell, subject to the boundary condition,
\bea
\label{bc}
\phi(x+b) = \phi(A^{-1} x)\;,
\eea
is well-defined on ${\mathscr M}$. These are the simplest examples of the rotated boundary conditions
of \cite{Nagy:2025qvg} where $b$ was taken in the temporal direction. 
Generalization to groups with two generators is straightforward, there will
be two independent boundary conditions corresponding to the two generators and the 4-torus is split up
into subvolumes both temporally and spatially.

The rotated boundary conditions are very much analogous to the shifted boundary conditions
\cite{Giusti:2010bb,Giusti:2011kt,Giusti:2012yj,Giusti:2014ila,Giusti:2016iqr}
introduced for
the study of a moving frame in QFT. In this case the velocity $v_i$ in Euclidean signature is purely imaginary,
$i v_i = T y_i$ and define a spatial vector $y$. The temporal boundary condition imposed on fields in the path integral becomes,
\bea
\phi(x + b) = \phi( x - y )\;,
\eea
where again $b$ is in the temporal direction while the shift by $y$ is spatial. 
Translation by $y$ is also an isometry of
course and (\ref{bc}) corresponds to replacing these translations with the other type of isometries,
rotations. This observation leads to the interpretation of QFT on some of the non-trivial Bieberbach manifolds
as describing a spatially rotating volume at finite temperature.

\section{Fourier transform}
\label{fouriertransform}

In any QFT application the discussion of 1-particle states will need the Fourier transform. 
In this section we discuss the Fourier transform on ${\mathscr M}$ and how it descends from ${\mathbb T}^4$.
The basis of the reciprocal lattice, $\Lambda^*$, will be denoted by $f^a$, so we have
$f^a\cdot e^b = \delta_{ab}$. Then momenta on ${\mathbb T}^4$ are labelled by integers as $p = 2\pi n_a
f^{a}$ as usual. Fourier modes $e^{i px}$ on ${\mathbb T}^4$ can be made well-defined on ${\mathscr M}$ by
making them invariant with respect to $G$,
\bea
\label{ff}
\phi_p(x) = \frac{1}{|G|} \sum_{(A,b)\in G} e^{i p (Ax+b)}\;.
\eea
However, not all modes are independent, 
if $q = A^Tp$, with some $(A,b)\in G$ then $\phi_p(x)$ and $\phi_q(x)$ are related by,
\bea
\phi_p(x) = e^{i p b} \phi_q(x)\;,
\eea
hence momenta $p$ and $Ap$ should be regarded as equivalent. Let us introduce $O_p$ for the orbit of $p$,
then all momenta on $O_p$ are equivalent.

Furthermore, some momenta lead to identically vanishing $\phi_p(x)$, which in 3-dimensional 
crystallography are usually called systematic absences or extinctions. We
will call these forbidden momenta. In 3-dimensions all groups are abelian, in 4-dimensions as we have seen
non-abelian groups are also allowed and the set of forbidden momenta is somewhat more complicated. In
order to describe them let us first inspect an abelian example, ${\mathbb Z}_k$. There is a single
generator $g=(A,b)$ and $g^k=1$ (up to lattice translations) 
which means $A$ is a rotation by $2\pi/k$ and the basis vectors $e^a$ can
be chosen such that $b = e^1/k$ and is a direction left invariant by $A$. 
If $A^T p = p$ i.e. the momenta are in a 2-dimensional plane left
invariant by the rotation, then, with $p = 2\pi n_a f^a$,
\bea
\label{abelian}
\phi_p(x) &=& \frac{e^{ipx}}{k} \sum_{(A,b)\in {\mathbb Z}_k} e^{ipb} = 
\frac{e^{ipx}}{k} \sum_{j=0}^{k-1} e^{2\pi i n_1 j/k}\;,
\eea
which is zero if $e^{2\pi i n_1/k}$ is not unity, i.e. if $n_1$ is not a multiple of $k$. Hence forbidden
momenta in these abelian cases are given by only 2 integers and a condition on one of them.

The analysis for the non-abelian groups follows the above with the only complication that momenta $p$
can have different little groups $G_p = \{ (A,b) \in G \; | \; A^Tp = p \}$. Let us introduce $O_p$ for
the orbit of $p$ as well, $O_p = \{ A^T p\; | \; (A,b) \in G \}$. Then, by collecting the 
Fourier modes with the same $x$-dependence in (\ref{ff}), we get,
\bea
\phi_p(x) = \frac{1}{|G|} \sum_{q \in O_p} c_p(q) e^{iqx}\;, \qquad \qquad \qquad \qquad 
c_p(q) = \sum_{\substack{(A,b) \in G \\ A^Tp = q}} e^{ipb}
\eea
which will be identically zero if for all $q \in O_p$ the coefficients are vanishing, $c_p(q)=0$. 
This condition is equivalent to the character sum of $G_p$ vanishing,
\bea
\label{cond}
\sum_{(A,b) \in G_p} e^{ipb} = 0\;,
\eea
which defines the full set of forbidden momenta $p$.

In the abelian $G = {\mathbb Z}_k$ case $G_p$ is either just the identity or the full $G$. If the former,
the above equation is never satisfied, if the latter, then $p$ has zero components in the plane of rotation
and we get a condition on the projection of $p$ on $b$, which gives back the result from (\ref{abelian}).

In the non-abelian cases one needs to classify all $G_p$ little groups and impose (\ref{cond}). The result
for all 26 Bieberbach manifolds is given in appendix \ref{forbiddenmomenta}. Hence we have a full
accounting of equivalent as well as forbidden momenta for all flat, compact, orientable ${\mathscr M}$,
allowing for the calculation of free eigenvalues and their multiplicities. These can be used for
the calculation of various (regularized) determinants.

\section{Rotating space-time}
\label{rotatingspacetime}

The first application, considered in \cite{Nagy:2025qvg}, is the Euclidean path integral formulation of a
space-time at finite temperature undergoing rotation with a constant angular velocity. 
In Euclidean signature, in order to have no sign problem, the angular velocity is purely imaginary. 

The spatial geometry may be chosen as a cylinder and the rotation is around the symmetry
axis. There is a priori no conceptual problem with such a setup and a path integral formulation is
straightforward, but from a practical point of view the cylinder
geometry complicates any non-perturbative lattice calculation, especially if the effect of rotation is
incorporated into a non-trivial metric in the rotating frame. The continuum geometry allows
for infinitesimal rotations but any discretization breaks it. Nevertheless valuable information can be
gained by numerical lattice simulations of this setup \cite{Yamamoto:2013zwa, Ambrus:2014uqa, Chernodub:2016kxh, Chernodub:2017ref, Braguta:2021jgn, Braguta:2023iyx, Braguta:2023yjn, Braguta:2023tqz}.

Another approach, introduced in \cite{Nagy:2025qvg} starts with a usual ${\mathbb T}^4$ geometry, which
is straightforward to implement in a lattice setup and
instead of modifying the metric with angular velocity dependent factors (as in the case of a rotating
frame), the action in the path integral stays the same as without rotations but the temporal boundary conditions
on the fields are modified. Analytic continuation of the velocity leads to a periodic angle variable, 
$\omega = - i \vartheta T$, and $\vartheta$ is not arbitrary but
can only take values $\vartheta = 2\pi/k$ with $k=2,3,4,6$. Discreteness is straightforward, a 4-torus cannot
be rotated by arbitrary angles. The rotated temporal boundary conditions required are precisely of the
type (\ref{bc}) with $b$ in the temporal direction and $A \in SO(4)$ a rotation by $2\pi/k$ in a spatial
plane. Hence the geometry is ${\mathscr M} = {\mathbb T}^4 / {\mathbb Z}_k$, a flat, compact, orientable
4-manifold and the classification in Table \ref{table} gives us all the options, including the only
choices $k=2,3,4,6$. 

Some of the non-abelian cases can also give rise to a rotating space-time interpretation. What is
required is that one of the generators should shift in a preferred direction (the temporal direction)
and $\Lambda$ should split into a spatial lattice, orthogonal to the temporal direction. 
The rotational part of the generator should rotate
spatially only, and the other generator should both rotate and shift in the spatial subspace only.
The non-abelian nature of $G$ will be due to the two 3-dimensional rotations not commuting.
These requirements are possible in 7 of the non-abelian cases and 3 of the $\mathbb Z_2 \times \mathbb
Z_2$ cases and the column $k$ in Table \ref{table}
indicates the allowed angles, $\vartheta = 2\pi/k$, or a dash if such a split is not possible, hence a
rotating space-time interpretation is not possible.

As a curiosity we mention that the last item of Table \ref{table}, min.44.2.1.1 which corresponds to
one of the $A_4$ cases, has the interesting property that the spatial rotations do not lead to a
preferred axis. All the spatial directions are symmetric. The basis vectors, without loss of generality, 
can be chosen as
$e^1=(L,0,0,0),\; e^2=(0,L,0,0),\; e^3=(0,0,L,0),\; e^4=(0,0,0,L_0)$ where $L$ and $L_0$ are the two free
parameters in the metric shown in the $n$ column of Table \ref{table}. This spatial symmetry holds only
for this one among those geometries where a rotating interpretation is possible. 
Clearly in the ${\mathbb Z}_k$ cases spatially there is a preferred direction, the
rotation axis, and also in all others, except min.44.2.1.1.

The main advantage of the above approach 
is that traditional lattice techniques are immediately available, one simply needs to
impose rotated temporal (or both temporal and spatial) boundary conditions on the fields. 

\section{Summary and outlook}
\label{summary}

In this paper novel space-time geometries were described which are flat, compact and orientable just as
${\mathbb T}^4$. There are only finitely many different topologies in each dimension and in 4-dimensions
there are 26 non-trivial options; all of them can be realized as ${\mathbb T}^4 / G$ with a finite
group $G$. The classification follows similar techniques as 3-dimensional
crystallography, in one dimension higher. A novel feature of 4 (and higher) dimensions is that
$G$ can be non-abelian.

We have discussed scalar fields in detail but there is no problem with introducing gauge fields, subject
to similar rotated boundary conditions. Fermion fields and a Dirac operator can only be defined if
${\mathscr M}$ is a spin manifold which is the case for 23 out of the 26 cases.

As a first application we have described a path integral formulation of a rotating spatial volume at
finite temperature. In a Euclidean setup, in which we are working, the constant angular velocity needs to
be imaginary, in order to avoid a sign problem. The resulting angular variables $\vartheta_j$, given by
$i \omega_j = \vartheta_j T$ can only be discrete and exactly correspond to 7 cases
with cyclic groups ${\mathbb Z}_{2,3,4,6}$ with periodic spatial boundary conditions. 
There are two different topologies for $k=2,3,4$ and a single
one for $k=6$. Infinitesimal rotations are of course not possible because of the 3-torus spatial volume,
but finite rotations are, which in a path integral formulation can be implemented by imposing spatially
rotated temporal boundary conditions, in other words by fields on ${\mathbb T}^4 / {\mathbb Z}_k$. We
believe this formulation of a rotating space-time lends itself in the most straightforward way to non-perturbative
lattice calculations which was the primary motivation. The simplicity lies in not having to deal with 
a curved space-time.

We would like to point out that a recent work \cite{Cork:2026fxx} 
in infinite volume, $S^1 \times \mathbb R^3$ introduced the
Nahm-transform for rotating calorons. A similar approach could in principle be extended to our compact
$\mathbb T^4 / G$ manifolds using the full 4-dimensional Nahm-transform.

\section*{Acknowledgements}

DN would like to thank Rafa\l$\,$ Lutowski for numerous very enlightening discussions.
This work was supported by NKFIH grants No. TKP2021-NKTA-64, NKKP Excellence No. 151482 and K-147396.

\appendix

\section{Generators}
\label{generators}

The explicit generators, one or two, depending on the group, are given in this section for all 
26 non-trivial groups. The generators $g=(A,b)$ are expressed in the notation (\ref{nr}) and the
integer matrices $N$ and rational vectors $r$ are assembled into $5 \times 5$ matrices
\bea
\lmatrix{ccccc}
    N & r \\
    0 & 1
\rmatrix
\eea
The labels are the CARAT labels \cite{carat} from crystallography.

\bigskip

min.18.1.1.1 \quad $\mathbb{Z}_2$
\[
\lmatrix{ccccc}
1&0&0&0&\frac{1}{2}\\
0&1&0&0&0\\
0&0&-1&0&0\\
0&0&0&-1&0\\
0&0&0&0&1
\rmatrix
\]

min.18.1.2.1 \quad $\mathbb{Z}_2$
\[
\lmatrix{ccccc}
1&0&0&0&\frac{1}{2}\\
0&-1&0&0&0\\
0&0&0&1&0\\
0&0&1&0&0\\
0&0&0&0&1
\rmatrix
\]

min.35.1.1.1 \quad $\mathbb{Z}_3$
\[
\lmatrix{ccccc}
0&1&0&0&0\\
-1&-1&0&0&0\\
0&0&1&0&\frac{1}{3}\\
0&0&0&1&0\\
0&0&0&0&1
\rmatrix
\]

min.35.1.2.1 \quad $\mathbb{Z}_3$
\[
\lmatrix{ccccc}
1&0&0&0&\frac{1}{3}\\
0&0&1&0&0\\
0&0&0&-1&0\\
0&-1&0&0&0\\
0&0&0&0&1
\rmatrix
\]

min.25.1.1.1 \quad $\mathbb{Z}_4$
\[
\lmatrix{ccccc}
0&-1&0&0&0\\
1&0&0&0&0\\
0&0&1&0&\frac{1}{4}\\
0&0&0&1&0\\
0&0&0&0&1
\rmatrix
\]

min.25.1.2.2 \quad $\mathbb{Z}_4$
\[
\lmatrix{ccccc}
1&0&0&0&\frac{1}{4}\\
0&0&0&-1&0\\
0&1&0&1&0\\
0&0&-1&1&0\\
0&0&0&0&1
\rmatrix
\]

group.70.1.1.1 \quad $\mathbb{Z}_6$
\[
\lmatrix{ccccc}
1&1&0&0&0\\
-1&0&0&0&0\\
0&0&1&0&\frac{1}{6}\\
0&0&0&1&0\\
0&0&0&0&1
\rmatrix
\]

min.22.1.1.11 \quad $\mathbb{Z}_2 \times \mathbb{Z}_2$
\[
\lmatrix{ccccc}
-1&0&0&0&0\\
0&1&0&0&\frac{1}{2}\\
0&0&-1&0&\frac{1}{2}\\
0&0&0&1&0\\
0&0&0&0&1
\rmatrix
\lmatrix{ccccc}
1&0&0&0&\frac{1}{2}\\
0&-1&0&0&0\\
0&0&-1&0&0\\
0&0&0&1&0\\
0&0&0&0&1
\rmatrix
\]

min.22.1.1.12 \quad $\mathbb{Z}_2 \times \mathbb{Z}_2$
\[
\lmatrix{ccccc}
-1&0&0&0&0\\
0&1&0&0&0\\
0&0&-1&0&0\\
0&0&0&1&\frac{1}{2}\\
0&0&0&0&1
\rmatrix
\lmatrix{ccccc}
1&0&0&0&\frac{1}{2}\\
0&-1&0&0&0\\
0&0&-1&0&0\\
0&0&0&1&0\\
0&0&0&0&1
\rmatrix
\]

min.22.1.1.13 \quad $\mathbb{Z}_2 \times \mathbb{Z}_2$
\[
\lmatrix{ccccc}
-1&0&0&0&0\\
0&1&0&0&\frac{1}{2}\\
0&0&-1&0&0\\
0&0&0&1&\frac{1}{2}\\
0&0&0&0&1
\rmatrix
\lmatrix{ccccc}
1&0&0&0&\frac{1}{2}\\
0&-1&0&0&0\\
0&0&-1&0&0\\
0&0&0&1&0\\
0&0&0&0&1
\rmatrix
\]

min.22.1.1.15 \quad $\mathbb{Z}_2 \times \mathbb{Z}_2$
\[
\lmatrix{ccccc}
-1&0&0&0&0\\
0&1&0&0&\frac{1}{2}\\
0&0&-1&0&\frac{1}{2}\\
0&0&0&1&\frac{1}{2}\\
0&0&0&0&1
\rmatrix
\lmatrix{ccccc}
1&0&0&0&\frac{1}{2}\\
0&-1&0&0&0\\
0&0&-1&0&0\\
0&0&0&1&0\\
0&0&0&0&1
\rmatrix
\]

min.22.1.2.7 \quad $\mathbb{Z}_2 \times \mathbb{Z}_2$
\[
\lmatrix{ccccc}
-1&0&0&0&\frac{1}{2}\\
0&1&0&0&\frac{1}{2}\\
0&0&0&-1&0\\
0&0&-1&0&0\\
0&0&0&0&1
\rmatrix
\lmatrix{ccccc}
-1&0&0&0&0\\
0&1&0&0&\frac{1}{2}\\
0&0&0&1&0\\
0&0&1&0&0\\
0&0&0&0&1
\rmatrix
\]

min.22.1.4.7 \quad $\mathbb{Z}_2 \times \mathbb{Z}_2$
\[
\lmatrix{ccccc}
1&0&0&0&\frac{1}{2}\\
0&-1&0&0&\frac{1}{2}\\
0&0&0&-1&0\\
0&0&-1&0&0\\
0&0&0&0&1
\rmatrix
\lmatrix{ccccc}
-1&0&0&0&0\\
0&-1&0&0&0\\
0&0&1&0&-\frac{1}{2}\\
0&0&0&1&\frac{1}{2}\\
0&0&0&0&1
\rmatrix
\]

min.22.1.5.3 \quad $\mathbb{Z}_2 \times \mathbb{Z}_2$
\[
\lmatrix{ccccc}
-1&0&0&0&\frac{1}{2}\\
0&1&1&0&-\frac{1}{2}\\
0&0&-1&0&0\\
0&0&-1&1&0\\
0&0&0&0&1
\rmatrix
\lmatrix{ccccc}
-1&0&0&0&0\\
0&0&0&-1&0\\
0&0&1&0&\frac{1}{2}\\
0&-1&0&0&0\\
0&0&0&0&1
\rmatrix
\]

min.22.1.8.7 \quad $\mathbb{Z}_2 \times \mathbb{Z}_2$
\[
\lmatrix{ccccc}
-1&0&0&0&\frac{1}{2}\\
0&1&0&0&-\frac{1}{2}\\
0&-1&0&-1&\frac{1}{2}\\
0&-1&-1&0&0\\
0&0&0&0&1
\rmatrix
\lmatrix{ccccc}
-1&0&0&0&0\\
0&0&1&-1&-\frac{1}{2}\\
0&1&0&1&0\\
0&0&0&1&\frac{1}{2}\\
0&0&0&0&1
\rmatrix
\]

min.22.1.12.6 \quad $\mathbb{Z}_2 \times \mathbb{Z}_2$
\[
\lmatrix{ccccc}
1&0&2&0&-\frac{1}{2}\\
-1&0&-1&-1&\frac{1}{2}\\
0&0&-1&0&0\\
-1&-1&-1&0&0\\
0&0&0&0&1
\rmatrix
\lmatrix{ccccc}
-1&0&0&-2&-\frac{1}{2}\\
1&0&1&1&0\\
1&1&0&1&0\\
0&0&0&1&\frac{1}{2}\\
0&0&0&0&1
\rmatrix
\]

min.38.1.1.1 \quad $S_3$
\[
\lmatrix{ccccc}
-1&-1&0&0&0\\
0&1&0&0&0\\
0&0&1&0&-\frac{1}{2}\\
0&0&0&-1&0\\
0&0&0&0&1
\rmatrix
\lmatrix{ccccc}
0&1&0&0&0\\
-1&-1&0&0&0\\
0&0&1&0&0\\
0&0&0&1&-\frac{1}{3}\\
0&0&0&0&1
\rmatrix
\]

min.38.2.1.1 \quad $S_3$
\[
\lmatrix{ccccc}
0&-1&0&0&0\\
-1&0&0&0&0\\
0&0&1&0&-\frac{1}{2}\\
0&0&0&-1&0\\
0&0&0&0&1
\rmatrix
\lmatrix{ccccc}
-1&-1&0&0&0\\
1&0&0&0&0\\
0&0&1&0&0\\
0&0&0&1&-\frac{1}{3}\\
0&0&0&0&1
\rmatrix
\]

min.38.2.3.1 \quad $S_3$
\[
\lmatrix{ccccc}
-1&0&0&0&0\\
0&0&1&0&\frac{1}{2}\\
0&1&0&0&\frac{1}{2}\\
0&0&0&1&-\frac{1}{2}\\
0&0&0&0&1
\rmatrix
\lmatrix{ccccc}
1&0&0&0&-\frac{1}{3}\\
0&0&0&-1&0\\
0&1&0&0&0\\
0&0&-1&0&0\\
0&0&0&0&1
\rmatrix
\]

min.29.1.1.13 \quad $D_4$
\[
\lmatrix{ccccc}
1&0&0&0&0\\
0&-1&0&0&0\\
0&0&1&0&\frac{1}{2}\\
0&0&0&-1&\frac{1}{4}\\
0&0&0&0&1
\rmatrix
\lmatrix{ccccc}
0&-1&0&0&0\\
-1&0&0&0&0\\
0&0&1&0&\frac{1}{2}\\
0&0&0&-1&0\\
0&0&0&0&1
\rmatrix
\]

min.29.1.1.14 \quad $D_4$
\[
\lmatrix{ccccc}
1&0&0&0&\frac{1}{2}\\
0&-1&0&0&0\\
0&0&1&0&0\\
0&0&0&-1&\frac{1}{4}\\
0&0&0&0&1
\rmatrix
\lmatrix{ccccc}
0&-1&0&0&0\\
-1&0&0&0&0\\
0&0&1&0&\frac{1}{2}\\
0&0&0&-1&0\\
0&0&0&0&1
\rmatrix
\]

min.29.1.1.15 \quad $D_4$
\[
\lmatrix{ccccc}
1&0&0&0&\frac{1}{2}\\
0&-1&0&0&0\\
0&0&1&0&\frac{1}{2}\\
0&0&0&-1&\frac{1}{4}\\
0&0&0&0&1
\rmatrix
\lmatrix{ccccc}
0&-1&0&0&0\\
-1&0&0&0&0\\
0&0&1&0&\frac{1}{2}\\
0&0&0&-1&0\\
0&0&0&0&1
\rmatrix
\]

min.29.1.2.7 \quad $D_4$
\[
\lmatrix{ccccc}
-1&0&0&0&\frac{1}{4}\\
0&0&0&-1&0\\
0&0&1&0&\frac{1}{2}\\
0&-1&0&0&0\\
0&0&0&0&1
\rmatrix
\lmatrix{ccccc}
-1&0&0&0&0\\
0&0&1&-1&-\frac{1}{2}\\
0&1&0&1&0\\
0&0&0&1&\frac{1}{2}\\
0&0&0&0&1
\rmatrix
\]

group.81.1.1.5 \quad $D_6$
\[
\lmatrix{ccccc}
-1&0&0&0&0\\
1&1&0&0&0\\
0&0&-1&0&0\\
0&0&0&1&\frac{1}{2}\\
0&0&0&0&1
\rmatrix
\lmatrix{ccccc}
1&1&0&0&0\\
-1&0&0&0&0\\
0&0&1&0&\frac{1}{6}\\
0&0&0&1&0\\
0&0&0&0&1
\rmatrix
\]

min.44.1.2.1 \quad $A_4$
\[
\lmatrix{ccccc}
0&-1&1&1&\frac{1}{3}\\
-1&-1&1&1&0\\
0&-1&1&0&\frac{1}{3}\\
0&-1&0&1&0\\
0&0&0&0&1
\rmatrix
\lmatrix{ccccc}
-1&0&0&0&\frac{5}{6}\\
0&1&-1&-1&\frac{1}{2}\\
0&0&0&-1&0\\
0&0&-1&0&0\\
0&0&0&0&1
\rmatrix
\]

min.44.2.1.1 \quad $A_4$
\[
\lmatrix{ccccc}
0&0&1&0&0\\
-1&0&0&0&0\\
0&-1&0&0&0\\
0&0&0&1&-\frac{1}{3}\\
0&0&0&0&1
\rmatrix
\lmatrix{ccccc}
-1&0&0&0&-\frac{1}{2}\\
0&1&0&0&-\frac{1}{2}\\
0&0&-1&0&0\\
0&0&0&1&0\\
0&0&0&0&1
\rmatrix
\]

\section{Forbidden momenta}
\label{forbiddenmomenta}

In this appendix the forbidden momenta are listed for all 26 non-trivial
groups, as described in section \ref{fouriertransform}. 
For each case we list its CARAT name from crystallography \cite{carat}, $G$, 
the little groups and the corresponding forbidden momenta.

\bigskip

min.18.1.1.1 \quad $\mathbb{Z}_2$
\[
\begin{array}{lll}
\mathbb{Z}_2: & (n_1,\, n_2,\, 0,\, 0) & n_1 \ \mathrm{odd}
\end{array}
\]

\bigskip

min.18.1.2.1 \quad $\mathbb{Z}_2$
\[
\begin{array}{lll}
\mathbb{Z}_2: & (n_1,\, 0,\, n_3,\, n_3) & n_1 \ \mathrm{odd}
\end{array}
\]

\bigskip

min.35.1.1.1 \quad $\mathbb{Z}_3$
\[
\begin{array}{lll}
\mathbb{Z}_3: & (0,\, 0,\, n_3,\, n_4) & n_3 \ \mathrm{not\ a\ multiple\ of}\ 3
\end{array}
\]

\bigskip

min.35.1.2.1 \quad $\mathbb{Z}_3$
\[
\begin{array}{lll}
\mathbb{Z}_3: & (n_1,\, -n_4,\, -n_4,\, n_4) & n_1 \ \mathrm{not\ a\ multiple\ of}\ 3
\end{array}
\]

\bigskip

min.25.1.1.1 \quad $\mathbb{Z}_4$
\[
\begin{array}{lll}
\mathbb{Z}_4: & (0,\, 0,\, n_3,\, n_4) & n_3 \ \mathrm{not\ a\ multiple\ of}\ 4
\end{array}
\]

\bigskip

min.25.1.2.2 \quad $\mathbb{Z}_4$
\[
\begin{array}{lll}
\mathbb{Z}_4: & (n_1,\, -n_4,\, -n_4,\, n_4) & n_1 \ \mathrm{not\ a\ multiple\ of}\ 4
\end{array}
\]

\bigskip

group.70.1.1.1 \quad $\mathbb{Z}_6$
\[
\begin{array}{lll}
\mathbb{Z}_6: & (0,\, 0,\, n_3,\, n_4) & n_3 \ \mathrm{not\ a\ multiple\ of}\ 6
\end{array}
\]

\bigskip

min.22.1.1.11 \quad $\mathbb{Z}_2 \times \mathbb{Z}_2$
\[
\begin{array}{lll}
\bZ_2: & (0,\, n_2,\, 0,\, n_4) & n_2 \ \mathrm{odd} \\
\bZ_2: & (n_1,\, 0,\, 0,\, n_4) & n_1 \ \mathrm{odd} \\
\bZ_2: & (0,\, 0,\, n_3,\, n_4) & n_3 \ \mathrm{odd}
\end{array}
\]

\bigskip

min.22.1.1.12 \quad $\mathbb{Z}_2 \times \mathbb{Z}_2$
\[
\begin{array}{lll}
\mathbb{Z}_2 \times \mathbb{Z}_2: & (0,\, 0,\, 0,\, n_4) & n_4 \ \mathrm{odd} \\
\bZ_2: & (0,\, n_2,\, 0,\, n_4) & n_4 \ \mathrm{odd} \\
\bZ_2: & (n_1,\, 0,\, 0,\, n_4) & n_1 \ \mathrm{odd} \\
\bZ_2: & (0,\, 0,\, n_3,\, n_4) & n_4 \ \mathrm{odd}
\end{array}
\]

\bigskip

min.22.1.1.13 \quad $\mathbb{Z}_2 \times \mathbb{Z}_2$
\[
\begin{array}{lll}
\mathbb{Z}_2 \times \mathbb{Z}_2: & (0,\, 0,\, 0,\, n_4) & n_4 \ \mathrm{odd} \\
\bZ_2: & (0,\, n_2,\, 0,\, n_4) & n_2 + n_4 \ \mathrm{odd} \\
\bZ_2: & (n_1,\, 0,\, 0,\, n_4) & n_1 \ \mathrm{odd} \\
\bZ_2: & (0,\, 0,\, n_3,\, n_4) & n_4 \ \mathrm{odd}
\end{array}
\]

\bigskip

min.22.1.1.15 \quad $\mathbb{Z}_2 \times \mathbb{Z}_2$
\[
\begin{array}{lll}
\mathbb{Z}_2 \times \mathbb{Z}_2: & (0,\, 0,\, 0,\, n_4) & n_4 \ \mathrm{odd} \\
\bZ_2: & (0,\, n_2,\, 0,\, n_4) & n_2 + n_4 \ \mathrm{odd} \\
\bZ_2: & (n_1,\, 0,\, 0,\, n_4) & n_1 \ \mathrm{odd} \\
\bZ_2: & (0,\, 0,\, n_3,\, n_4) & n_3 + n_4 \ \mathrm{odd}
\end{array}
\]

\bigskip

min.22.1.2.7 \quad $\mathbb{Z}_2 \times \mathbb{Z}_2$
\[
\begin{array}{lll}
\mathbb{Z}_2 \times \mathbb{Z}_2: & (0,\, n_2,\, 0,\, 0) & n_2 \ \mathrm{odd} \\
\bZ_2: & (0,\, n_2,\, -n_4,\, n_4) & n_2 \ \mathrm{odd} \\
\bZ_2: & (0,\, n_2,\, n_3,\, n_3) & n_2 \ \mathrm{odd} \\
\bZ_2: & (n_1,\, n_2,\, 0,\, 0) & n_1 \ \mathrm{odd}
\end{array}
\]

\bigskip

min.22.1.4.7 \quad $\mathbb{Z}_2 \times \mathbb{Z}_2$
\[
\begin{array}{lll}
\bZ_2: & (n_1,\, 0,\, -n_4,\, n_4) & n_1 \ \mathrm{odd} \\
\bZ_2: & (0,\, 0,\, n_3,\, n_4) & n_3 + n_4 \ \mathrm{odd} \\
\bZ_2: & (0,\, n_2,\, -n_4,\, n_4) & n_2 \ \mathrm{odd}
\end{array}
\]

\bigskip

min.22.1.5.3 \quad $\mathbb{Z}_2 \times \mathbb{Z}_2$
\[
\begin{array}{lll}
\mathbb{Z}_2 \times \mathbb{Z}_2: & (0,\, -n_4,\, -n_4,\, n_4) & n_4 \ \mathrm{odd} \\
\bZ_2: & (0,\, n_2 + 2n_3,\, n_3,\, n_2) & n_2 \ \mathrm{odd} \\
\bZ_2: & (0,\, -n_4,\, n_3,\, n_4) & n_3 \ \mathrm{odd} \\
\bZ_2: & (n_1,\, -n_4,\, -n_4,\, n_4) & n_1 \ \mathrm{odd}
\end{array}
\]

\bigskip

min.22.1.8.7 \quad $\mathbb{Z}_2 \times \mathbb{Z}_2$
\[
\begin{array}{lll}
\bZ_2: & (0,\, n_2,\, -n_4,\, n_4) & n_2 + n_4 \ \mathrm{odd} \\
\bZ_2: & (0,\, n_2,\, n_2,\, n_4) & n_2 + n_4 \ \mathrm{odd} \\
\bZ_2: & (n_1,\, -n_4,\, -n_4,\, n_4) & n_1 \ \mathrm{odd}
\end{array}
\]

\bigskip

min.22.1.12.6 \quad $\mathbb{Z}_2 \times \mathbb{Z}_2$
\[
\begin{array}{lll}
\bZ_2: & (n_1,\, -n_4,\, n_1,\, n_4) & n_1 + n_4 \ \mathrm{odd} \\
\bZ_2: & (n_1,\, n_1,\, n_1,\, n_4) & n_1 + n_4 \ \mathrm{odd} \\
\bZ_2: & (n_1,\, n_1,\, -n_4,\, n_4) & n_1 + n_4 \ \mathrm{odd}
\end{array}
\]

\bigskip

min.38.1.1.1 \quad $S_3$
\[
\begin{array}{lll}
S_3: & (0,\, 0,\, n_3,\, 0) & n_3 \ \mathrm{odd} \\
\bZ_3: & (0,\, 0,\, n_3,\, n_4) & n_4 \ \mathrm{not\ a\ multiple\ of}\ 3 \\
\bZ_2: & (0,\, n_2,\, n_3,\, 0) & n_3 \ \mathrm{odd} \\
\bZ_2: & (n_1,\, 0,\, n_3,\, 0) & n_3 \ \mathrm{odd} \\
\bZ_2: & (n_1,\, n_1,\, n_3,\, 0) & n_3 \ \mathrm{odd}
\end{array}
\]

\bigskip

min.38.2.1.1 \quad $S_3$
\[
\begin{array}{lll}
S_3: & (0,\, 0,\, n_3,\, 0) & n_3 \ \mathrm{odd} \\
\bZ_3: & (0,\, 0,\, n_3,\, n_4) & n_4 \ \mathrm{not\ a\ multiple\ of}\ 3 \\
\bZ_2: & (-n_2,\, n_2,\, n_3,\, 0) & n_3 \ \mathrm{odd} \\
\bZ_2: & (n_1,\, 2n_1,\, n_3,\, 0) & n_3 \ \mathrm{odd} \\
\bZ_2: & (2n_2,\, n_2,\, n_3,\, 0) & n_3 \ \mathrm{odd}
\end{array}
\]

\bigskip

min.38.2.3.1 \quad $S_3$
\[
\begin{array}{lll}
S_3: & (0,\, -n_4,\, -n_4,\, n_4) & n_4 \ \mathrm{odd} \\
\bZ_3: & (n_1,\, -n_4,\, -n_4,\, n_4) & n_1 \ \mathrm{not\ a\ multiple\ of}\ 3 \\
\bZ_2: & (0,\, n_2,\, n_2,\, n_4) & n_4 \ \mathrm{odd} \\
\bZ_2: & (0,\, n_2,\, -n_4,\, n_4) & n_2 \ \mathrm{odd} \\
\bZ_2: & (0,\, -n_4,\, n_3,\, n_4) & n_3 \ \mathrm{odd}
\end{array}
\]

\bigskip

min.29.1.1.13 \quad $D_4$
\[
\begin{array}{lll}
D_4: & (0,\, 0,\, n_3,\, 0) & n_3 \ \mathrm{odd} \\
\bZ_4: & (0,\, 0,\, n_3,\, n_4) & n_4 \ \mathrm{not\ a\ multiple\ of}\ 4 \\
\bZ_2: & (n_1,\, 0,\, n_3,\, 0) & n_3 \ \mathrm{odd} \\
\bZ_2: & (-n_2,\, n_2,\, n_3,\, 0) & n_3 \ \mathrm{odd} \\
\bZ_2: & (n_1,\, n_1,\, n_3,\, 0) & n_3 \ \mathrm{odd} \\
\bZ_2: & (0,\, n_2,\, n_3,\, 0) & n_3 \ \mathrm{odd}
\end{array}
\]

\bigskip

min.29.1.1.14 \quad $D_4$
\[
\begin{array}{lll}
D_4: & (0,\, 0,\, n_3,\, 0) & n_3 \ \mathrm{odd} \\
\bZ_4: & (0,\, 0,\, n_3,\, n_4) & 2 n_3 + n_4 \ \mathrm{not\ a\ multiple\ of}\ 4 \\
\bZ_2: & (n_1,\, 0,\, n_3,\, 0) & n_1 \ \mathrm{odd} \\
\bZ_2: & (-n_2,\, n_2,\, n_3,\, 0) & n_3 \ \mathrm{odd} \\
\bZ_2: & (n_1,\, n_1,\, n_3,\, 0) & n_3 \ \mathrm{odd} \\
\bZ_2: & (0,\, n_2,\, n_3,\, 0) & n_2 \ \mathrm{odd}
\end{array}
\]

\bigskip

min.29.1.1.15 \quad $D_4$
\[
\begin{array}{lll}
D_4: & (0,\, 0,\, n_3,\, 0) & n_3 \ \mathrm{odd} \\
\bZ_4: & (0,\, 0,\, n_3,\, n_4) & n_4 \ \mathrm{not\ a\ multiple\ of}\ 4 \\
\bZ_2: & (n_1,\, 0,\, n_3,\, 0) & n_1 + n_3 \ \mathrm{odd} \\
\bZ_2: & (-n_2,\, n_2,\, n_3,\, 0) & n_3 \ \mathrm{odd} \\
\bZ_2: & (n_1,\, n_1,\, n_3,\, 0) & n_3 \ \mathrm{odd} \\
\bZ_2: & (0,\, n_2,\, n_3,\, 0) & n_2 + n_3 \ \mathrm{odd}
\end{array}
\]

\bigskip

min.29.1.2.7 \quad $D_4$
\[
\begin{array}{lll}
D_4: & (0,\, -n_4,\, -n_4,\, n_4) & n_4 \ \mathrm{odd} \\
\bZ_4: & (n_1,\, -n_4,\, -n_4,\, n_4) & n_1 + 2 n_4 \ \mathrm{not\ a\ multiple\ of}\ 4 \\
\bZ_2: & (0,\, -n_4,\, n_3,\, n_4) & n_3 \ \mathrm{odd} \\
\bZ_2: & (0,\, n_2,\, n_2,\, n_4) & n_2 + n_4 \ \mathrm{odd} \\
\bZ_2: & (0,\, n_2,\, -n_4,\, n_4) & n_2 + n_4 \ \mathrm{odd} \\
\bZ_2: & (0,\, n_2 + 2n_3,\, n_3,\, n_2) & n_2 \ \mathrm{odd}
\end{array}
\]

\bigskip

group.81.1.1.5 \quad $D_6$
\[
\begin{array}{lll}
D_6: & (0,\, 0,\, 0,\, n_4) & n_4 \ \mathrm{odd} \\
\bZ_6: & (0,\, 0,\, n_3,\, n_4) & n_3 \ \mathrm{not\ a\ multiple\ of}\ 6 \\
\bZ_2: & (n_1,\, 2n_1,\, 0,\, n_4) & n_4 \ \mathrm{odd} \\
\bZ_2: & (0,\, n_2,\, 0,\, n_4) & n_4 \ \mathrm{odd} \\
\bZ_2: & (n_1,\, n_1,\, 0,\, n_4) & n_4 \ \mathrm{odd} \\
\bZ_2: & (-n_2,\, n_2,\, 0,\, n_4) & n_4 \ \mathrm{odd} \\
\bZ_2: & (2n_2,\, n_2,\, 0,\, n_4) & n_4 \ \mathrm{odd} \\
\bZ_2: & (n_1,\, 0,\, 0,\, n_4) & n_4 \ \mathrm{odd}
\end{array}
\]

\bigskip

min.44.1.2.1 \quad $A_4$
\[
\begin{array}{lll}
A_4: & (0,\, 0,\, -n_4,\, n_4) & n_4 \ \mathrm{not\ a\ multiple\ of}\ 3 \\
\bZ_3: & (n_1 + n_4,\, -n_1 - n_4,\, n_1,\, n_4) & n_1 - n_4 \ \mathrm{not\ a\ multiple\ of}\ 3 \\
\bZ_3: & (n_1 - n_3,\, n_1 - n_3,\, n_3,\, -3n_1 + 2n_3) & n_3 \ \mathrm{not\ a\ multiple\ of}\ 3 \\
\bZ_3: & (-n_2 - n_4,\, n_2 + n_4,\, n_2,\, n_4) & n_2 - n_4 \ \mathrm{not\ a\ multiple\ of}\ 3 \\
\bZ_3: & (-n_3 - n_4,\, -n_3 - n_4,\, n_3,\, n_4) & n_3 - n_4 \ \mathrm{not\ a\ multiple\ of}\ 3 \\
\bZ_2: & (0,\, -n_3 - n_4,\, n_3,\, n_4) & n_3 + n_4 \ \mathrm{odd} \\
\bZ_2: & (-n_3 - n_4,\, 0,\, n_3,\, n_4) & n_3 + n_4 \ \mathrm{odd} \\
\bZ_2: & (0,\, 0,\, n_3,\, n_4) & n_3 + n_4 \ \mathrm{odd}
\end{array}
\]

\bigskip

min.44.2.1.1 \quad $A_4$
\[
\begin{array}{lll}
A_4: & (0,\, 0,\, 0,\, n_4) & n_4 \ \mathrm{not\ a\ multiple\ of}\ 3 \\
\bZ_3: & (n_1,\, -n_1,\, n_1,\, n_4) & n_4 \ \mathrm{not\ a\ multiple\ of}\ 3 \\
\bZ_3: & (-n_3,\, -n_3,\, n_3,\, n_4) & n_4 \ \mathrm{not\ a\ multiple\ of}\ 3 \\
\bZ_3: & (-n_2,\, n_2,\, n_2,\, n_4) & n_4 \ \mathrm{not\ a\ multiple\ of}\ 3 \\
\bZ_3: & (n_1,\, n_1,\, n_1,\, n_4) & n_4 \ \mathrm{not\ a\ multiple\ of}\ 3 \\
\bZ_2: & (0,\, n_2,\, 0,\, n_4) & n_2 \ \mathrm{odd} \\
\bZ_2: & (n_1,\, 0,\, 0,\, n_4) & n_1 \ \mathrm{odd} \\
\bZ_2: & (0,\, 0,\, n_3,\, n_4) & n_3 \ \mathrm{odd}
\end{array}
\]

\printcredits

\end{document}